# Three unsolved problems in physics of Pc1 magnetospheric waves

**A. Guglielmi**

*Institute of Physics of the Earth of the Russian Academy of Science, Moscow, Russia*

E-mail: guglielmi@mail.ru

**Abstract**

In this paper we analyzed the open problems in the physics of ultra-low-frequency electromagnetic waves Pc1, which also known as the "pearls" (frequency range 0.2 - 5 Hz). A number of theoretical problems arose because the standard model of excitation and propagation of waves in the magnetosphere, the foundations of which were laid half a century ago, is unable to explain observable properties of Pc1. In this article the problems are formulated and discussed in context of the general physics of geoelectromagnetic waves. In addition, one problem of experimental character is considered. The essence of the problem is that the observations testify in favor of the idea of some human impact on the Pc1 oscillation mode, but difficulties in interpreting of observations indicate the need for further experimental research.

**Contents**

# 1. Introduction

Ultra low frequency Pc1 electromagnetic waves, known as the *pearls* or *pearl necklace*, are excited sporadically in the magnetosphere, or rather in the outer radiation belt in the frequency range of 0.2 - 5 Hz [Troitskaya, Guglielmi, 1967; Kangas et al., 1998]. Waves were discovered by E. Sucksdorff and L. Harang at Sodankylä and Tromsø observatories, respectively [Sucksdorff, 1936; Harang, 1936]. At the EGU-2006 General Assembly in Vienna a special session titled "Pc1 Pearl Waves: Discovery, Morphology and Physics" devoted to the 70[th] anniversary of this discovery was held by the initiative of J. Kangas and the author. The session aroused wide interest, and this is understandable since the Pc1 waves play an important role in the solar-terrestrial relations [Guglielmi, Kangas, 2007].

The Pc1 wave theory is based on the achievements of magnetohydrodynamics [Alfven, 1950] and the general theory of electromagnetic waves in plasma [Ginzburg, 1962]. Early history of the theoretical interpretation of experimental facts relates to the first half of 60s of the last century. In this period K. Yanagihara, L. Tepley, and J. Cornwall suggested principal elements of the so-called standard model of Pc1 excitation and propagation [Yanagihara, 1963; Cornwall, 1965; Tepley, 1965]. In brief, the model is as follows (see for example [Kangas et al., 1998]).

Pc1 are the electromagnetic waves that occur sporadically in the outer radiation belt as a result of ion-cyclotron instability. In the magnetosphere Pc1 emission propagate along the geomagnetic field lines in the form of Alfven waves. They penetrate through the ionosphere to the Earth's surface. Alfven waves are partially reflected from the ionosphere in magneto-conjugated areas. They come back into the radiation belt that leads it to self-excitation. At ionospheric heights the Alfven waves are partially transformed into magnetosound waves which propagate along the Earth's surface over long distances in the so-called ionospheric MHD waveguide. The dominant factors in the excitation and propagation of Pc1 are the impact of the solar wind on the magnetosphere and the impact of solar ultraviolet and X-ray radiation on the ionosphere.

The standard model of Pc1 is still used with some variations. Meanwhile, by the end of the 60s years of the last century three problems of theoretical character were found [Guglielmi, 1967, 1971] (see also [Guglielmi, Troitskaya, 1973; Guglielmi, Potapov, 2012]). It turned out that the standard model is unable to explain the stable discreteness of Pc1, which manifests itself in the fact that Pc1 have the form of a quasi-periodic sequence of wave packets. The problem of discreteness relates closely with still unresolved question on the mode of self-excitation of the radiation belt. The quasi-linear theory of ion-cyclotron instability predicts the soft self-excitation regime. However the discreteness of Pc1 indicates indirectly the hard self-oscillation regime.

The phenomenological models of hard self-excitation, that was proposed so far, are not quite satisfactory.

The third problem is the need to take into account the multi-ion composition of the magnetospheric plasma. Satellite observations suggest that the magnetosphere contains protons and alpha particles of solar origin, and oxygen ions of ionospheric origin. The first specific feature of this problem is that the Alfven wave in multi-component plasma must cross a number of bands of the opacity. The second peculiarity is related to the theoretical prediction concerning the existence of ion-cyclotron resonator; its role in the formation of the Pc1 spectrum has not been established yet.

Generally speaking, the emergence of problems in the process of research indicates the normal development of our understanding of the waves in magnetosphere. However, since the problems remain open for so long, it is advisable to draw a special attention to them. Long existence of unresolved problems is undesirable as it throws down a challenge of our ability to understand the physics of electromagnetic waves of natural origin. In this paper we analyze these theoretical problems.

## 2. The problem of self-excitation

We would like to discuss the problem of self-excitation in the framework of the phenomenological Landau's theory. Let us introduce the order parameter $\varepsilon$ and the control parameter $\Lambda$. The master equation has the form

$$\frac{d\varepsilon}{dt} = 2\Gamma\varepsilon \tag{1}$$

Here the nonlinear growth rate $\Gamma(\Lambda,\varepsilon)$ equals

$$\Gamma(\Lambda,\varepsilon) = \gamma(\Lambda) - \alpha\varepsilon, \tag{2}$$

where $\alpha$ is the Landau's constant (see [Landau, Lifshitz, 1987]). The linear growth rate $\gamma(\Lambda)$ near the threshold of self-excitation is proportional to the control parameter:

$$\gamma(\Lambda) = \eta(\Lambda - \Lambda_c) \tag{3}$$

Here $\Lambda_c$ is the critical value of $\Lambda$, and $\eta$ is the proportionality factor.

With reference to our problem it is naturally to choose the parameter $\varepsilon(t)$ characterizing the current state of the oscillatory system to be equal to the square of the Pc1 amplitude averaged over the period of oscillations: $\varepsilon = <E^2>$. The theory of ion-cyclotron instability of the radiation belt [Cornwall, 1965] prompts selection of the control parameter: $\Lambda = N(T_\perp/T_\| - 1)$. Here $N$ is the concentration of the energetic protons, $T_\|$ and $T_\perp$ are the longitudinal and transversal temperatures of these protons, respectively. Hence, the theory contains three phenomenological parameters – $\eta$, $\Lambda_c$ and $\alpha$. The linear theory of waves in plasma is used to calculate the parameters $\eta$ and $\Lambda_c$; Landau's constant $\alpha$ is usually calculated in the framework of quasi-linear theory (e.g., [Kangas et al., 1998]).

In equilibrium $d\varepsilon/dt = 0$. According to (1) the equilibrium can be disordered, if $\varepsilon = 0$, or ordered if $\varepsilon \neq 0$, but $\Gamma(\varepsilon, \Lambda) = 0$. The sign of $\alpha$ in the expression (2) for the nonlinear growth rate is very important. If $\alpha > 0$ ($\alpha < 0$), then we have a dynamic system with a soft (hard) self-excitation of oscillations.

Let us consider firstly the case of $\alpha > 0$. In the subcritical state ($\Lambda < \Lambda_c$) the system is stable and there are no oscillations ($E = 0$). When crossing threshold ($\Lambda > \Lambda_c$) the magnetosphere (or, more precisely, some oscillatory subsystem of the magnetosphere) makes phase transition of the second order. The amplitude of oscillation is proportional to the square root of supercriticality, $E \propto \sqrt{\Lambda - \Lambda_c}$. Thus, the system with a soft mode of self-excitation is characterized by disordered state when $\Lambda < \Lambda_c$, and ordered state when $\Lambda > \Lambda_c$. When $\Lambda > \Lambda_c$, the oscillations are excited under the influence of an arbitrarily small perturbation.

The case of $\alpha < 0$ is more interesting in the context of this work. In contrast to the dynamic system with a soft self-excitation, the system with a hard self-excitation is metastable in the subcritical state. In other words, the system is stable with respect to infinitesimal perturbation, but it can switch into a self-oscillating mode under the influence of the trigger, i.e. perturbation of small but finite amplitude.

For the analysis of excitation of a metastable system we should replace (2) with the more general formula

$$\Gamma(\Lambda, \varepsilon) = \gamma(\Lambda) - \alpha\varepsilon - \beta\varepsilon^2, \qquad (4)$$

where $\beta$ is one more phenomenological parameter of the theory, $\beta > 0$ [Landau, Lifshitz, 1987]. Taking into account (1), (3), and (4) we find that the system is stable at $\Lambda < \Lambda'_c$, where

$$\Lambda'_c = \Lambda - \frac{\alpha^2}{2\eta\beta} \qquad (5)$$

At $\Lambda = \Lambda_c$ the system abruptly switches to self-oscillation with the amplitude $E = \sqrt{|\alpha|/\beta}$. In the range of $\Lambda'_c < \Lambda < \Lambda_c$, the system is metastable, but it can go into the oscillatory mode under the action of exogenous trigger with the amplitude $E > \sqrt{|\alpha|/2\beta}$.

The question about hard self-excitation of the radiation belt was raised in [Guglielmi, 1971; Guglielmi, Troitskaya, 1973] in relation with the problem of discreteness of Pc1 waves. The difficulty consists in the following. The quasi-linear theory of the wave-particle interaction, widely adopted in plasma physics, predicts the positive value of the Landau's constant $\alpha$. In other words, the theory predicts a soft mode of self-excitation of the radiation belt. However, when $\alpha > 0$ then it is difficult to explain not only the discreteness of Pc1, but the clear signs of influence of the exogenous triggers on the mode of oscillation. Thus, the problem of self-excitation remains open.

### 3. The problem of discreteness

Usually Pc1 oscillations occur spontaneously and stop without a visible reason about an hour after the start. The most important property of such a séance, or better to say, a series of oscillations was noted by E. Sucksdorff in his pioneering work [Sucksdorff, 1936]. He drew attention to the fact that each series consists of a quasi-periodic sequence of the wave packets. That is why he called the series of oscillations as *pearl necklace*.

Repetition period of the wave packet is approximately two minutes. In the geometrical optics approximation, the repetition period is equal to the travel time of the Alfven waves along geomagnetic field line from one magneto-conjugated point to another and back. With this in mind, K. Yanagihara made appropriate observation and found that the "pearls" appear alternately at the conjugate points [Yanagihara, 1963]. Independent observations have confirmed this result (see the reviews [Troitskaya, Guglielmi, 1967; Kangas et al., 1998]). This was the basis for creation of the standard model briefly described in the Introduction.

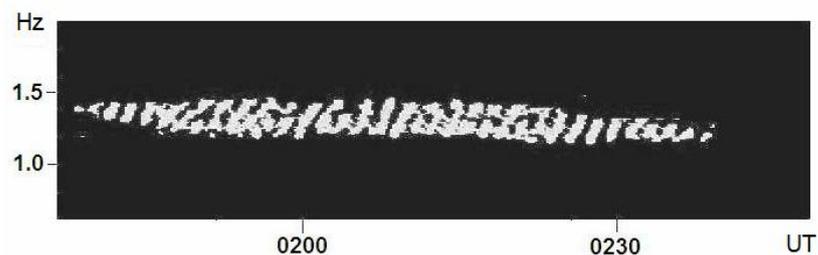

**Fig. 1.** Dynamical spectrun of Pc1 electromagnetic oscillations (GO Borok, 29.10.1984).

Fig. 1 shows the dynamical spectrum of Pc1 oscillations registered in obs. Borok. We see that the series of Pc1 lasts about one hour and consists of the discrete signals with increasing frequency. At one time the idea of a single package of Alfven waves was popular. It was assumed that the packet oscillates along the geomagnetic field lines from one hemisphere to another, losing energy upon reflection from the ionosphere, but making up for the loss due to resonant interaction with energetic proton of the radiation belt in equatorial zone of the magnetosphere. The frequency growth in separate "pearls" was explained by mechanism of gyrofrequency dispersion, especially appreciable near the equator of a wave trajectory. The duration of a series was explained by the gradual exhaustion of a free energy in autonomous closed oscillatory system.

It seems amazing that a long time geophysicists not be concerned about the problem arising in relation with the discreteness of Pc1. Meanwhile, the discreteness, which is reproduced inevitably from case to case, is incompatible with the notion of a linear wave packet propagating in a dispersive medium. Besides, in the linear approach it is impossible to understand a stable balance between the losses of wave energy in the ionosphere and the pumping of wave energy in the radiation belt. Balance problem was solved in the framework of the quasi-linear theory of plasma instabilities. But it does not address the issue of discreteness for the reason that the quasi-linear theory predicts $\alpha > 0$, i.e. the soft self-excitation regime.

These reasoning led the author of this paper to an attempt to explain the discreteness, rejecting notions of the closeness of oscillatory system, while maintaining other elements of the standard model of Pc1. Prototype for a modified model was the so-called continuous-flow cultivator, well known in microbiology. As a result, the model of the Alfven continuous-flow resonator was proposed [Guglielmi, 1971]. The idea is that the active filling enters into the resonator through Eastern wall of the ray tube in the form of energetic protons drifting along the azimuth in the geomagnetic trap (see the monograph [Guglielmi, Troitskaya, 1973] for details). The model of Alfven continuous-flow resonator was picked up [Tagirov et al., 1986; Trakhtengerts et al., 1986; Trakhtengerts, Demekhov, 2002]. However, contrary to expectations, it did not solve the discreteness problem. First, all versions of the theory were phenomenological which in itself is not very satisfactory. Second, over time a new circumstance came to light [Guglielmi et al., 2000, 2001], which called into question the applicability of the standard model at all, but this will be discussed in the next section.

### 4. The problem of heavy ions

To explain the essence of the third problem we recall some formulas from the theory of interaction between the Alfven waves and protons of the radiation belt. The condition of resonant exchange of energy between the wave with a frequency $\omega$ and the proton moving with the velocity $v_\parallel$ along the magnetic field lines has the form

$$\omega = \Omega_p - k_\parallel v_\parallel, \tag{6}$$

where $\Omega_p = eB/m_p c$ is the proton gyrofrequency, $k_\| = (\omega/c)n_\|$ is the longitudinal component of the wave vector **k**, $B$ is the value of magnetic field **B**, $e$ and $m_p$ are the charge and mass of the proton, $c$ is the velocity of light. If the vectors **k** and **B** are parallel, then $n_\| = \sqrt{\varepsilon_\perp + g}$; if **k** and **B** are almost perpendicular to each other, then $n_\| \approx \sqrt{\varepsilon_\perp}$. Here $\varepsilon_\perp$ and $g$ are the known components of the dielectric permeability tensor.

If the plasma contains only electrons and protons, as it is supposed in the standard model, then

$$\varepsilon_\perp = \frac{(c/c_A)^2}{1-(\omega/\Omega_p)^2}, \quad g = \left(\frac{\omega}{\Omega_p}\right)\varepsilon_\perp, \tag{7}$$

where $c_A$ is the Alfven velocity. The formula $k_\| = (\omega/c)n_\|$ with taking into account of (7) indicates that the band of transparency for Alfven waves is bounded from above by the gyrofrequency of protons. If $\omega > \Omega_p$, the waves do not propagate. But if $\omega < \Omega_p$ at the top of the wave trajectory, this inequality is not violated throughout the propagation path from one conjugate point in the ionosphere to the other. Hence, inequality $\omega < \Omega_p$ is an important condition for the applicability of the standard model of Pc1.

However, we must take into consideration that in addition to protons the real magnetosphere contains appreciable amounts of ions $O^+$ and $He^+$ of the ionospheric origin, and ions $He^{++}$ of the solar origin. These ions are often called "heavy" because for them charge to mass ratio is less than for the ion $H^+$. Here we will also use this term although it is not quite strict. Now, instead of (7) it is necessary to use the formulae

$$\varepsilon_\perp = \sum_{e,i} \frac{\omega_0^2}{\Omega^2 - \omega^2}, \quad g = \sum_{e,i} \frac{\Omega \omega_0^2}{\omega(\Omega^2 - \omega^2)}, \tag{8}$$

where the summation is over all species of charged particles, $\omega_0 = (4\pi e^2 N/m)^{1/2}$ is the plasma frequency, $\Omega = eB/mc$ is the gyrofrequency, $e$, $m$ and $N$ are the charge, mass and concentration of particles of a given specie. Inequality $\omega < \Omega_p$, referred to above, should be replaced by more stringent inequality $\omega < \Omega_{O^+}$. The wave propagates freely from one conjugate point to another if only $\omega < \Omega_{O^+}$ throughout the propagation path.

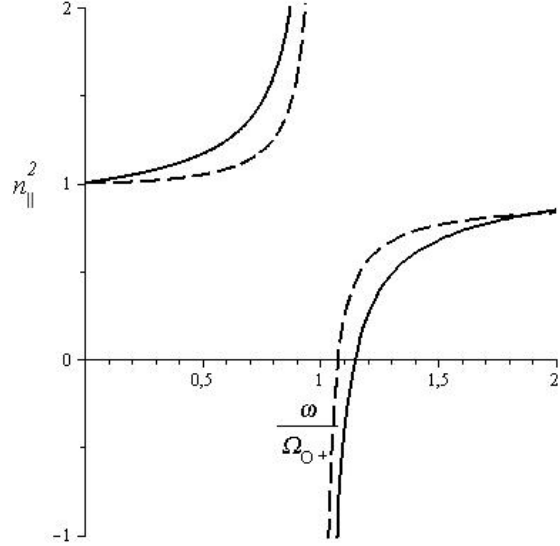

**Fig.2.** Frequency dependence of the squared refraction index for hydrogen plasma with a small admixture of $O^+$ ions at quasi-parallel (solid line) and quasi-transverse (dashed line) propagation. The squared refraction index is expressed in units of $c^2/c_A^2$.

However, the condition $\omega < \Omega_{O^+}$ is wittingly violated in the vicinity of equator of the Pc1 wave trajectory. This is evidenced by evaluation of the resonance frequency by the formula (6) [Cornwall, 1965], as well as the satellite observations of Pc1 in the magnetosphere (e.g., see [Kangas et al., 1998]. But if $\omega > \Omega_{O^+}$ at the equator of trajectory, then the character of wave propagation changes radically [Guglielmi et al., 2000, 2001; Guglielmi, Potapov, 2012]. Fig. 2 gives an idea of the dispersion curves in the hydrogen plasma in presence of a small admixture of single-charged oxygen ions. The ion concentrations equal $N_{H^+} = 100$ cm$^{-3}$ and $N_{O^+} = 1$ cm$^{-3}$. We see that both for longitudinal and for quasi-transverse propagation there exist a pole of $n_\parallel^2(\omega)$ at the gyrofrequency of $O^+$ ions, and zero at the so-called cutoff frequency. The band of opacity ($n_\parallel^2 < 0$) is located between the zero and pole. Qualitatively, this picture does not changed if we replace $O^+$, for example, by $He^+$ [Guglielmi, 1967].

Let us assume that the wave propagates from the equator to the North, and let $\omega > \Omega_{O^+}$ on the equator, where the magnetic field is minimal. At the equator, the wave frequency exceeds also the cutoff frequency, since otherwise the wave propagation is impossible. The cutoff frequency is attained at some distance from the equator. Here $k_\parallel = 0$, and the reflection of wave occurs if the opacity bandwidth is large enough. (The band of opacity is wider, if the concentration of heavy ions is higher.) The same occurs in a case when the wave propagates from the equator to the South. Thus, the theory predicts that a peculiar ion-cyclotron resonator there exists in the magnetosphere [Guglielmi et al., 2000]. (When using more exact expressions, it would be necessary to speak about a set of ion-cyclotron resonators, but here we would not stop on it.) Characteristic size of cavity along geomagnetic field lines is much smaller than the length of these lines. In other words, the resonator is located high above the ionosphere in a rather narrow equatorial zone.

The existence of ion-cyclotron resonator follows directly from the multicomponent composition of the magnetospheric plasma. It points to the need of complete revision of the standard Pc1 model. The new problems are added to the problems that were considered above. One of them is related to the question about the mechanism of wave leakage from the resonator to the earth's surface.

## 5. Discussion

We have considered three theoretical problems of the physics of the Pc1 magnetospheric waves. In this section, we will point to one more non-trivial problem. In contrast to the considered problems it has experimental character. We are talking about weak but statistically significant anthropogenic impact on the Pc1 oscillation mode, synchronized by the precise time signals. On experience effects of both excitation and suppression of Pc1 under the action of anthropogenic triggers are observed. The problem is the difficulty to interpret the observations. This indicates a need for additional pilot study.

The question of pulsed influence of the industrial activity on Pc1 synchronized by signals of Universal Time was put in the paper [Guglielmi et al., 1978]. In recent years it has been discussed in detail in the literature (e.g., see [Guglielmi, Zotov, 2012; Zotov et al., 2013]). Therefore, referring the reader to these publications, we confine ourselves here to two remarks.

First, there is no doubt that the triggers are electromagnetic, rather than acoustic pulses from the technosphere. This is evidenced by a small delay between the time supposed sending a pulse and the sudden change in Pc1 oscillation mode. Secondly, the sudden suppression of Pc1 oscillations under the action of trigger is especially characteristic for systems with hard self-excitation.

But how can technosphere consumption energy be synchronized by the precise time signals? What are the specific sources of anthropogenic triggers, and what is a mechanism of their effects on oscillatory systems of the magnetosphere? These are the open questions which else it is necessary to answer.

# 6. Conclusion

We have considered three topical problems of the theory of Pc1 magnetospheric waves, and pointed one problem of experimental character. In conclusion, we state these problems in the form of brief questions:

1. What is the mode of the Pc1 self-excitation in the radiation belt, soft or hard?
2. What is the mechanism of spontaneous transformation of the wave field to the observed Pc1 waves in the form of quasi-periodic sequence of *pearls*?
3. Are there any ion-cyclotron resonators in the equatorial zone of the radiation belt? If yes, what is their role in the formation of the experimentally observed Pc1 dynamic spectrum ?
4. What is the mechanism of Pc1 excitation under the action of electromagnetic pulses of industrial origin synchronized by the marks of Universal Time?

These issues are interrelated. More than that, the answer to any of them will help to resolve other issues. For example, the creation of a realistic model of hard self-excitation of the radiation belt would make clearer the effect of Pc1 synchronization by hour markers, and positive answer to question of the existence of ion-cyclotron resonators allow to elaborate approaches to the problems of discreteness and self-excitation.

*Acknowledgments*. I thank B.V. Dovbnya, B. Klain, A. Potapov, L. Sobisevich, A. Sobisevich and O. Zotov for many helpful discussions and critical remarks. This work was supported by the RFBR (13-05-00066, 13-05-00529), and Program 4 of Basic Research of the RAS Presidium (Project 6.2).